%% file: ms-astero-spin-orbit.tex
\documentclass[fleqn,usenatbib]{mnras}

\usepackage{newtxtext,newtxmath}
\usepackage[switch]{lineno}
\usepackage{multirow}
\usepackage[T1]{fontenc}

\DeclareRobustCommand{\VAN}[3]{#2}
\let\VANthebibliography\thebibliography
\def\thebibliography{\DeclareRobustCommand{\VAN}[3]{##3}\VANthebibliography}

\usepackage{graphicx}	
\usepackage{amsmath}	
\usepackage{xcolor}

\newcommand{\st}[1]{_\mathrm{#1}}
\newcommand{\yr}{\,\mathrm{yr}}

\title[Spin-orbit alignments from \emph{Kepler} and \emph{Gaia}]
{Projected spin-orbit alignments from \emph{Kepler} asteroseismology and \emph{Gaia} astrometry}

\author[W. H. Ball et al.]{
Warrick H. Ball,\thanks{E-mail: W.H.Ball@bham.ac.uk}
Amaury H.M.J. Triaud,
Emily Hatt,
Martin B. Nielsen
and William J. Chaplin
\\
School of Physics \& Astronomy, University of Birmingham, Edgbaston, Birmingham B15 2TT, United Kingdom
}

\date{Accepted 2023 January 23. Received 2023 January 20; in original form 2022 October 12}

\pubyear{2023}

\begin{document}
\label{firstpage}
\pagerange{\pageref{firstpage}--\pageref{lastpage}}
\maketitle

\begin{abstract}
The angle between the rotation and orbital axes of stars 
in binary systems---the obliquity---is an important indicator 
of how these systems form and evolve but few such measurements exist.
We combine the sample of astrometric orbital inclinations from \emph{Gaia} DR3
with a
sample of solar-like oscillators in which rotational inclinations have been measured
using asteroseismology.
We supplement our sample with one binary whose
visual orbit has been determined using speckle interferometry
and present the projected spin-orbit alignments in five systems.
We find that each system, and the overall sample, is consistent with alignment
but there are important caveats.
First, the asteroseismic rotational
inclinations are fundamentally ambiguous and, second, we can only measure the
projected (rather than true) obliquity.
If rotational and orbital inclinations are independent and isotropically-distributed,
however, the likelihood of drawing our data by chance is less than a few per cent.
Though small, our data set argues against uniformly random obliquities in binary systems.
We speculate that dozens more measurements could be made using data from NASA's \emph{TESS} mission, mostly in red giants.
ESA's \emph{PLATO} mission will likely produce hundreds more spin-orbit measurements
in systems with main-sequence and subgiant stars.
\end{abstract}

\begin{keywords}
asteroseismology -- binaries: general -- stars: rotation
\vspace{-0.15em} 
\end{keywords}


\section{Introduction} \label{s:intro}

Obliquity is the angle between the rotation spin vector of a celestial object and its orbital spin vector.
The Sun's obliquity, for example, is about $7^\circ$ relative to the invariable plane of the Solar system. 
Na\"ively, we expect orbital and rotational spins to be aligned because a collapsing cloud imparts its angular momentum to a protostar and its protoplanetary disc, but few measurements exist to confirm this.
Most measurements have been obtained for transiting exoplanets via the Rossiter--McLaughlin effect \citep{Rossiter24,McLaughlin24}. 

A wide range of spin-orbit angles have been measured, with some planets in polar and retrograde orbits \citep{Triaud18,Albrecht22}. It is, however, unclear whether the
systems arrived in these configurations immediately after they formed or whether they evolved into them later.  Recently, \citet{Christian22} showed that binary companions are preferentially inclined relative to transiting planets, and proposed that the binary companion might align the protoplanetary disk,
thereby influencing how planets form.
If some planetary systems can be inclined \citep[e.g.][]{Hjorth21} and have a stellar companion, one should expect spin-orbit alignment as well as spin-orbit misalignment in binary systems.  

\citet{Hale94} studied 86 stars in 73 binary and higher-order systems and concluded that binary stars with orbital separations $a<30~\rm AU$ are aligned, while misalignment is common in more widely separated systems. 
\citet{Justesen20} re-analysed Hale's sample and found it insufficient to make any statements about the distribution of spin-orbit angle with orbital separation. Most recent work producing quality spin-orbit measurements has focused on short-period binaries, typically eclipsing, by modelling the Rossiter--McLaughlin effect \citep[e.g.][]{Kopal42,Gimenez06}. This includes high-mass eclipsing binaries such as the inclined DI Herculis \citep{Albrecht09}, and inclined CV Velorum \citep{Albrecht14}, or low-mass eclipsing binaries such as EBLM J1219-39 \citep{Triaud13} and EBLM J0608-59 \citep{Kunovac20}, both of which show alignment. 

Instead of the Rossiter-McLaughlin effect, \citet{Marcussen22} used apsidal motion to infer obliquity, and found that only 3 out of 51 surveyed binaries have spin-orbit misalignment but most were also short period binaries (the longest is $\sim 100~\rm days$). Unfortunately, close binary stars are a tricky sample to handle, since tidal interactions are expected to realign the rotation and orbital spins. 

More evidence is clearly necessary, particularly for binary systems with separations between $1$ and $50~\rm AU$, which have typically been harder to probe because eclipses are less likely and it is difficult to schedule observations with which to model the Rossiter--McLaughlin effect. 
Here we show how to measure the projected spin-orbit angle for non-eclipsing binaries that are spatially resolved or not, by combining asteroseismic measurements of the stellar inclination $i\st{rot}$ with astrometric measurements of the orbital inclination $i\st{orb}$. We use \emph{Kepler} results for the asteroseismology \citep{hall2021}, and \emph{Gaia} Data Release 3 (DR3) for the orbital parameters. At the moment only four measurements are possible but we expect several dozens might eventually be produced once all the \emph{Gaia} data is released, and thanks to new measurements of $i\st{rot}$ to be produced from \emph{TESS} and \emph{PLATO} 
asteroseismic data. 
We also include literature orbital data for one system where \emph{Gaia} provides a
spectroscopic solution but not an astrometric one.
Our approach is similar to those of \citet{LeBouquin09} and \citet{Sahlmann11}. 

\section{Methods}
\label{s:methods}

Most main-sequence solar-like oscillators rotate slowly,
in which case a star's pulsations can be described by spherical
harmonics, characterised by their angular degree $\ell$ and azimuthal order $m$,
multiplied by a radial eigenfunction, characterised by a radial order $n$.  
For each $\ell$, there are $2\ell+1$
azimuthal orders $-\ell\leq m\leq\ell$ that pulsate at the same frequency
if the star is spherically symmetric.  Slow rotation with period $P\st{rot}$
perturbs the frequencies by approximately $m/P\st{rot}$, lifting the degeneracy
such that modes of a given $n$ and $l$ form multiplets.\footnote{This is mathematically
the same as Zeeman splitting, where the presence of a magnetic field breaks the degeneracy
between electron orbitals, which are also described by spherical harmonics.}
This is known as \emph{rotational splitting}.

Under the standard assumption of energy equipartition between the components
of a rotationally-split multiplet, the relative amplitudes of the components
depend on the inclination angle of the rotation axis $i\st{rot}$ \citep{gizon2003}.
For example, modes with $\ell=1$ and $m\pm1$ are almost invisible if a star
is seen pole on, leaving only the $m=0$ component visible.  Conversely,
if a star is seen edge on, the $m=0$ component is almost invisible and only
the $m=\pm1$ pair is clearly observed.
Thus, one can in principle measure $i\st{rot}$ from high-quality observations
of solar-like oscillators, and this method has been widely applied to data
from \emph{Kepler}.
For example, \citet{campante2016} measured the stellar inclination angles
of 25 solar-like oscillators that host transiting planets
and concluded that the systems are all consistent with alignment.

\citet{hall2021} created a hierarchical Bayesian model to fit the mode frequencies,
including rotational inclination angles, of 91 stars observed by \emph{Kepler}
during its nominal mission.  This is the largest sample of asteroseismic
rotation inclinations available for main-sequence solar-like oscillators
and the one we selected to compare with the \emph{Gaia} measurements.
We used the \url{gaia-kepler.fun} cross-match to determine the corresponding \emph{Gaia} DR3
source IDs and queried the \texttt{gaiadr3.nss\_two\_body\_orbit} table for each star.

Four stars have astrometric solutions---KICs 4914923, 6933899, 9025370 and 12317678---,
for which we calculated the orbital inclinations
using the method of \citet{binnendijk1960} as described by \citet{halbwachs2022}.
In short, the orbital elements of the system are given in the \emph{Gaia}
data in terms of the Thiele--Innes elements, \citep{thiele1883,vandenbos1926}
\begin{align}
A&=\phantom{-}a (\cos\omega\cos\Omega - \sin\omega\sin\Omega\cos i\st{orb}) \\
B&=\phantom{-}a (\cos\omega\sin\Omega + \sin\omega\cos\Omega\cos i\st{orb}) \\
F&=-a (\sin\omega\cos\Omega + \cos\omega\sin\Omega\cos i\st{orb}) \\
G&=-a (\sin\omega\sin\Omega - \cos\omega\cos\Omega\cos i\st{orb})
\end{align}
where $a$ is the semi-major axis, $\omega$ the argument of periastron and $\Omega$
the longitude of the ascending node. i.e., $a$, $\omega$, $\Omega$ and $i\st{orb}$
are the Campbell elements.  \citet{halbwachs2022} give formulae to convert the Thiele--Innes elements
to the Campbell elements.
Uncertainties are propagated by drawing a sample of $10^5$ points from a normal
distribution with the mean and covariance given by the \emph{Gaia} data.  We report
the means and standard deviations of the derived samples in Table~\ref{t:data}.
The secondary mass $M_2$ is derived from the astrometric mass function 
under the assumption that the secondary is much fainter than the primary
\citep[see eq.~(15) of ][]{halbwachs2022}.  The primary's mass is taken from
\citet{hall2021}.

KIC~7510397 is an exception.  The \emph{Gaia} data
only contains an entry for the system as a single-lined spectroscopic binary so
we have used the orbital parameters from \citet{appourchaux2015}, who
analysed the solar-like oscillations detected in both stars in the binary.  
Their results included an orbital fit to speckle interferometry
that extended previous data presented by \citet{horch2012}
and we have included this value in Table~\ref{t:data}.

\section{Results} 
\label{s:data}

Fig.~\ref{f:spin-orbit} shows a comparison of the inclination angles of the rotational and orbital axes.
The asteroseismic measurement
cannot distinguish between angle $i\st{rot}$ and $180^\circ-i\st{rot}$,
so both are shown.  
The measurement for each system is what we would observe if they were aligned.
i.e., alignment is not ruled out in any of the five systems.
We cannot, however, conclude that the systems are truly aligned
but return to the significance of our result in Sec.~\ref{s:discuss}.

Three further stars are identified in the cross-match as spectroscopic binaries with
measured orbital periods: KICs 7206837, 7510397 and 9098294.  The reported orbital periods
for KICs 7206837 and 9098294 are consistent with the asteroseismically-measured rotation rates.
It is unclear if this implies tidal-locking with a close companion or that the rotation rate
has been mistaken for an orbital period.  
The orbital period for KIC 7510397 in the \emph{Gaia}
data of $61.63\pm0.47\,\mathrm{d}$ differs significantly from the period of $13.8^{+0.6}_{-0.5}\yr$ given by \citet{appourchaux2015},
though the \emph{Gaia} eccentricity of $0.515\pm0.029$ is only mildly inconsistent ($\lesssim 2\sigma$)
with their value of 
$0.583^{+0.016}_{-0.025}$.

Two further stars---KICs 1435467 and 8379927---have measured radial-velocity trends
but not complete orbital solutions.  They appear in the \texttt{gaiadr3.nss\_non\_linear\_spectro}
table but not \texttt{gaiadr3.nss\_two\_body\_orbit}. KIC 8379927 is known to be a spectroscopic
binary with an orbital period of about $4.8\,\mathrm{yr}$ \citep{griffin2007}.
With more data, \emph{Gaia} might determine a spectroscopic orbit and perhaps astrometric solution
for this system.

Pertinent data from both data sets for all seven stars with non-single solutions are listed in 
Table~\ref{t:data}, supplemented by the orbital inclination of KIC 7510397 
by \citet{appourchaux2015}.
We note that KIC 6933899 has a relatively long and very eccentric orbit, with $P\st{orb}=11.13\pm1.25\yr$ and
$e=0.917\pm0.008$. The orbital period is several times longer than the 34-month duration of the \emph{Gaia} data in DR3
but the goodness-of-fit statistic ($1.5$) and significance of the parameters suggest that this is a genuine
solution.

\section{Discussion and conclusion}
\label{s:discuss}

We have presented here only a preliminary study of the projected spin-orbit alignments
in systems with rotation inclinations measured through asteroseismology 
and orbital inclinations through astrometry. The sample of astrometric solutions
will only increase as \emph{Gaia} steadily takes more data but the results are
already significant.

\begin{figure}
	\includegraphics[width=\columnwidth]{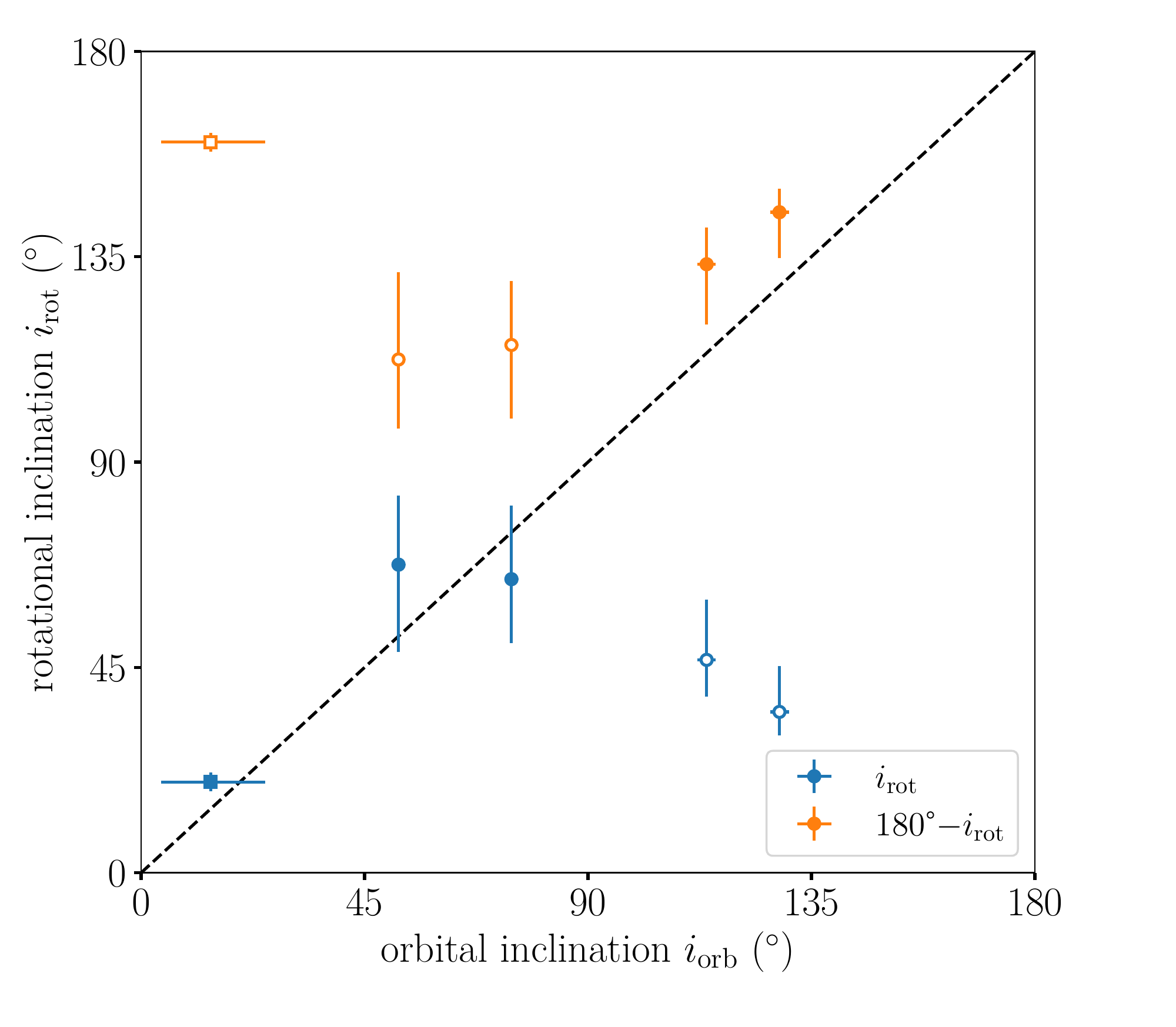}
	\caption{Comparison of rotation inclination angles by \citet{hall2021} and
    orbital inclination angles from \emph{Gaia} (circles) or, for KIC~7510397 
    \citep[][square]{appourchaux2015}.  The rotation angles $i\st{rot}$
	are ambiguous, so both $i\st{rot}$ and $180^\circ-i\st{rot}$ are plotted.
	The filled points indicate which value of $i\st{rot}$ is closer to the one-to-one line.}
    \label{f:spin-orbit}
\end{figure}

For each of the five systems in Fig.~\ref{f:spin-orbit}, the measurements
of $i\st{rot}$ and $i\st{orb}$ are what one would expect if they were aligned
but we cannot rule out misalignment for two reasons.
First, the asteroseismic measurement of $i\st{rot}$ cannot distinguish between
a measurement $i\st{rot}$ and $180^\circ-i\st{rot}$ so, far from $i\st{rot}=90^\circ$,
each system is roughly as likely to be aligned as misaligned.
Second, the true obliquity $\psi$ can lie anywhere between
$|i\st{rot}-i\st{orb}|$ and $i\st{rot}+i\st{orb}$.
True alignment can only be confirmed using the projected obliquity
if $i\st{rot}=i\st{orb}=0^\circ$.

\begin{figure}
	\includegraphics[width=\columnwidth]{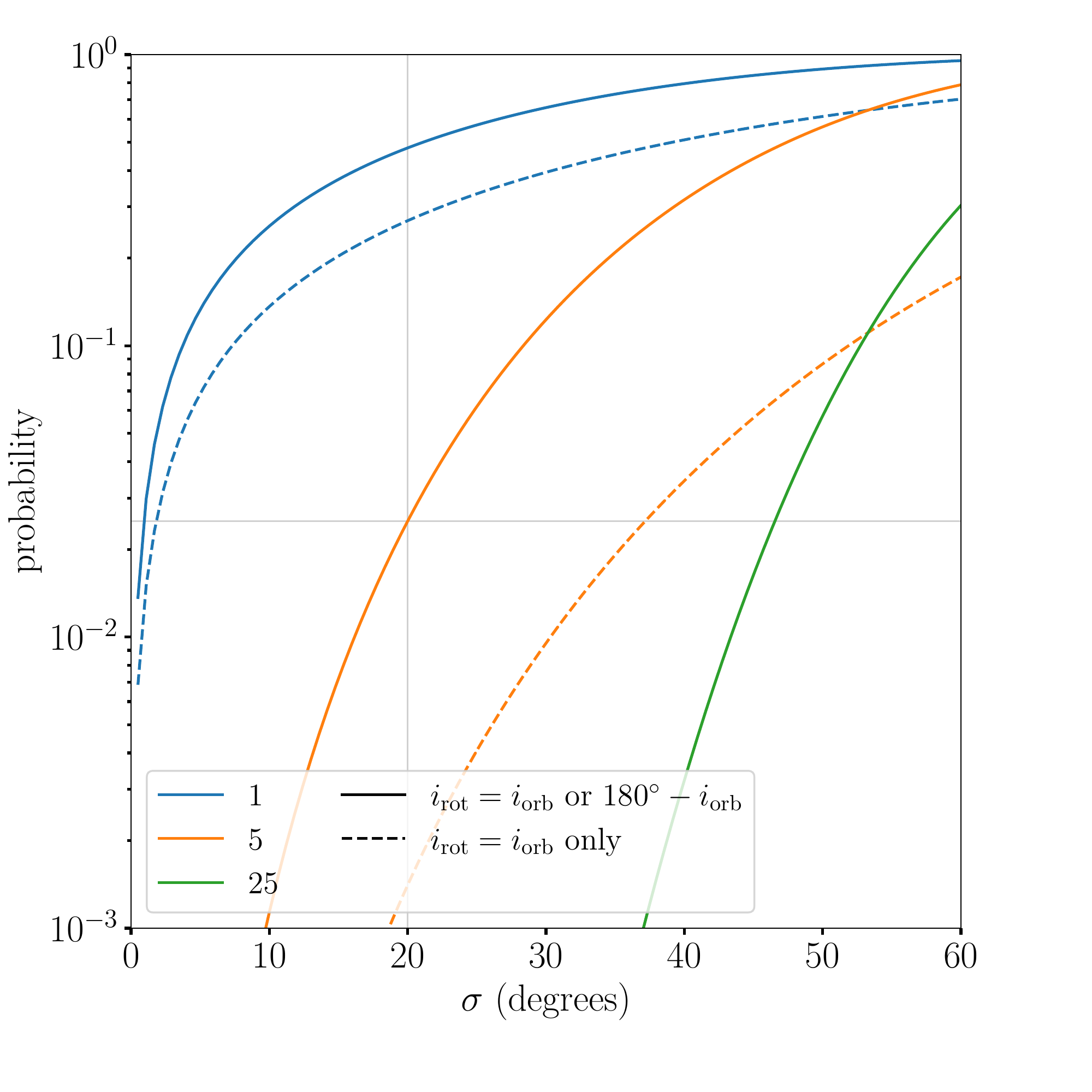}
	\caption{Probability of a sample of stars with isotropically-distributed and independent
 rotation inclinations $i\st{rot}$ and orbital inclinations $i\st{orb}$
 all having $i\st{rot}$ within $\sigma$ of $i\st{orb}$ (dashed lines) or within $\sigma$ 
 of either $i\st{orb}$ or $180^\circ-i\st{orb}$ (solid lines).
 The blue, orange and green curves show samples of $1$, $5$ or $25$ stars,
 respectively.  The solid grey lines show where $\sigma=20^\circ$; the corresponding
 probability is 2.5 per cent.}
    \label{f:significance}
\end{figure}

The fact that our results do not rule out alignment in any system
is nevertheless striking and we can compute the probability of measuring such
data---five data points consistent 
with $i\st{rot}=i\st{orb}$ or $180^\circ-i\st{orb}$---under the assumption 
that rotational and inclination axes are distributed isotropically and independently.  
In this case, the underlying joint distribution of $i\st{rot}$ and $i\st{orb}$ is
$\propto\sin i\st{rot}\sin i\st{orb}$.
For simplicity, we have integrated the region where $i\st{rot}$
is within some range $\pm\sigma$ of either $i\st{orb}$ or
$180^\circ-i\st{orb}$.
Fig.~\ref{f:significance} shows the likelihood of finding
different numbers of stars in this region, as a function of
the parameter $\sigma$.
The curve for $5$ stars corresponds to our sample.
If we take $\sigma=20^\circ$ as a representative width for the observed uncertainties, the likelihood
of measuring our data in a population of isotropically-distributed inclinations
is about 2.5 per cent.
Although we cannot conclude that all the systems in our sample are aligned,
our data is significantly at odds with the assumption of isotropic and independent rotation and orbital
inclinations.

Measuring rotational inclinations with asteroseismology has so far required space-based photometry.
Most of these results, including those used here, employ data from \emph{Kepler} 
but similar measurements have been made using data from CoRoT \citep[e.g.~HD~52265,][]{gizon2013}.
Aside from a sufficient signal-to-noise ratio to detect the solar-like oscillations, 
asteroseismic measurements of rotation
benefit from time-series that are several times longer than the stellar rotation rate,
which is several weeks for Sun-like stars.

NASA's \emph{Transiting Exoplanet Survey Satellite} (\emph{TESS}) is observing most of the sky
in 27.4-day-long sectors. Though most targets are only observed in a few sectors
separated by long gaps, some stars are in regions of the sky that \emph{TESS} observes
continuously for up to 13 sectors (roughly one year).  Solar-like oscillators
in these regions could potentially have their rotational inclination angles
measured, which could increase the sample of spin-orbit measurements.

\begin{figure}
	\includegraphics[width=\columnwidth]{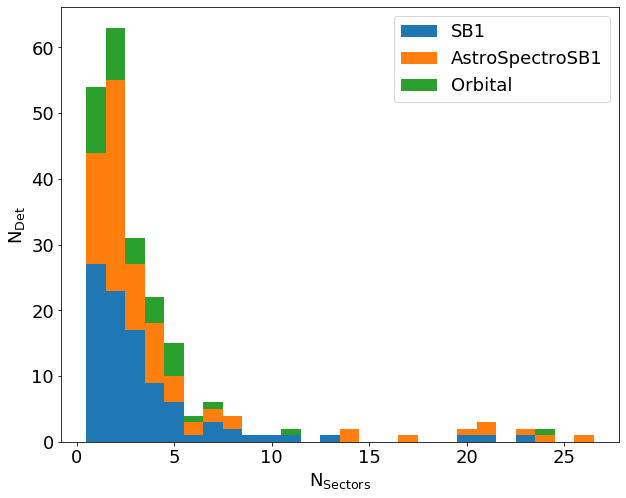}
	\caption{Stacked histogram showing the number of stars with orbital solutions 
	from \emph{Gaia} and solar-like oscillations
	  in their short-cadence \emph{TESS} light curves, as a function of the number of sectors
	  of \emph{TESS} data.}
    \label{f:hist}
\end{figure}

To explore \emph{TESS}'s potential further, 
we cross-matched the table of \emph{Gaia} two-body orbits with the catalogue by \citet{hatt2023}
of stars showing solar-like oscillations in short-cadence \emph{TESS} data.  
Fig.~\ref{f:hist} shows a histogram of the number of stars in the \emph{Gaia} table
that also show solar-like oscillations in their short-cadence light curves,
as a function of the number of sectors in which \emph{TESS} has observed them.  
There are 26 solar-like oscillators with astrometric orbits and
at least four sectors of \emph{TESS} data, though most of these are giants.  
Of these stars, six have $\log g > 3.4$, compared to the minimum $\log g = 3.91$ in the sample of \citet{hall2021}.
The rotational inclinations of red giants certainly can be measured---\citet{gehan2021}
have measured 1139 in data from \emph{Kepler}---but we have not analysed any here.

Finally, ESA's upcoming \emph{PLATO} mission \citep{rauer2014} will measure solar-like
oscillations in thousands of cool subgiants and main-sequence stars.  Even if we only
assume a yield of about 5 per cent, like the sample presented here, then \emph{PLATO}'s core
sample of $\sim15\,000$ stars would add hundreds of spin-orbit measurements.
We re-iterate that the sample is currently limited by the number of orbital
inclinations that have been measured, which can only increase as \emph{Gaia} continues
its observations.
Our results thus demonstrate the enormous potential to assemble a large
sample of projected spin-orbit angles, with which we will
be able to investigate the dynamics of binary stars.

\begin{table*}
\renewcommand{\arraystretch}{1.2} 
\input{table.tex}
\caption{Table of pertinent properties for the stars that appear both in the asteroseismic sample by \citet{hall2021}
  and \emph{Gaia}'s tables of non-single stars.  The first four stars are shown in Fig.~\ref{f:spin-orbit},
  as is KIC 7510397, for which we list the orbital inclination from \citet{appourchaux2015}.
  Symmetric uncertainties are indicated in brackets for that many final digits of the relevant number.}
  \label{t:data}
\end{table*}

\section*{Acknowledgements}

The authors would like to thank Oliver Hall for insightful discussions. 
WHB and WJC thank the UK Science and Technology Facilities Council (STFC) for support under grant ST/R0023297/1.  EJH and MBN acknowledge the support of STFC.
This research is in part funded by the European Union's Horizon 2020 research and innovation programme (grants agreements n$^{\circ}$ 803193/BEBOP). 
This work has made use of data from the European Space Agency (ESA) mission
\emph{Gaia} (\url{https://www.cosmos.esa.int/gaia}), processed by the \emph{Gaia}
Data Processing and Analysis Consortium (DPAC,
\url{https://www.cosmos.esa.int/web/gaia/dpac/consortium}). Funding for the DPAC
has been provided by national institutions, in particular the institutions
participating in the \emph{Gaia} Multilateral Agreement.


This work made use of the \emph{Gaia}--\emph{Kepler} crossmatch database created by Megan Bedell.\footnote{\url{https://gaia-kepler.fun}}

\section*{Data Availability}

Data from \emph{Gaia} is publicly available from the \emph{Gaia} Archive.\footnote{\url{https://gea.esac.esa.int/archive/}}
Python scripts to recreate Figs~\ref{f:spin-orbit} 
and \ref{f:significance}
and the data of which part is presented
in Table~\ref{t:data} are available from a public repository.\footnote{\url{https://gitlab.com/warrickball/kepler-gaia-spin-orbit}}
Other data underlying this article will be shared on reasonable request to the corresponding author.



\bibliographystyle{mnras}
\bibliography{ms-astero-spin-orbit} 







\bsp	
\label{lastpage}
\end{document}

%% file: table.tex
\begin{tabular}{cccccccccc}
\hline
Solution type & KIC & $i\st{orb}/{}^\circ$ & $i\st{rot}/{}^\circ$ & $P\st{orb}/\mathrm{d}$ & $P\st{rot}/\mathrm{d}$ & $e$ & $a/\mathrm{AU}$ & $M_1/\mathrm{M}_\odot$ & $M_2/\mathrm{M}_\odot$ \\ 
\hline
\multirow{2}{*}{AstroSpectroSB1} & 4914923 & $113.9 \pm 1.8$ & $46.6_{-8.1}^{+13.3}$ & $99.2443(664)$ & $21.40_{-3.53}^{+5.39}$ & $0.212(9)$ & $1.301(31)$ & $1.06_{-0.05}^{+0.06}$ & $0.51$\\ 
& 6933899 & $74.6 \pm 0.6$ & $64.3_{-14.0}^{+16.1}$ & $4065(455)$ & $28.91_{-4.83}^{+3.70}$ & $0.917(8)$ & $12.15(88)$ & $1.13_{-0.03}^{+0.03}$ & $0.56$\\ 
\hline
\multirow{2}{*}{Orbital} & 9025370 & $51.8 \pm 1.0$ & $67.5_{-19.1}^{+15.2}$ & $239.124(454)$ & $24.71_{-4.67}^{+3.67}$ & $0.271(28)$ & $1.417(22)$ & $0.97_{-0.03}^{+0.03}$ & $0.18$\\ 
& 12317678 & $128.6 \pm 1.9$ & $35.3_{-5.2}^{+10.1}$ & $80.8435(599)$ & $5.20_{-0.92}^{+1.80}$ & $0.393(36)$ & $1.019(30)$ & $1.34_{-0.01}^{+0.04}$ & $0.68$ \\
\hline
& 7206837 &  & $31.7_{-2.8}^{+3.2}$ & $4.05012(7)$ & $3.97_{-0.45}^{+0.55}$ & $0.002(10)$ & & $1.30_{-0.03}^{+0.03}$ & \\ 
SB1 & 7510397 & $\left[14^{+11}_{-10}\right]$ & $19.9_{-2.0}^{+2.0}$ & $61.6302(4664)$ & $6.13_{-0.64}^{+0.67}$ & $0.515(29)$ & & $1.37_{-0.02}^{+0.02}$ & \\ 
& 9098294 &  & $58.2_{-16.3}^{+21.0}$ & $20.1013(26)$ & $27.21_{-7.00}^{+5.75}$ & $0.018(10)$ & & $0.97_{-0.03}^{+0.02}$ & \\ 
\hline
Radial velocity & 1435467 &  & $63.4_{-6.6}^{+10.2}$ & & $6.54_{-0.62}^{+0.76}$ & & & $1.32_{-0.05}^{+0.03}$ & \\ 
trend & 8379927 &  & $63.3_{-2.3}^{+2.5}$ & & $9.20_{-0.23}^{+0.25}$ & & & $1.12_{-0.04}^{+0.04}$ & \\ 
\hline
\end{tabular}